\documentclass{kluwer}    
\usepackage{epsfig}

\newdisplay{guess}{Conjecture}

\begin{document}                                                                                   
\begin{article}
\begin{opening}         
\title{
Energetics of jets from X-ray binaries
\footnote{To be published in Proc. 3rd microquasar workshop, Granada,
Eds. A. J. Castro-Tirado, J. Greiner and J. M. Paredes, Astrophysics and
Space Science}
} 
\author{Rob \surname{Fender}}  
\runningauthor{Rob Fender}
\runningtitle{Jet Energetics}
\institute{Astronomical Institute `Anton Pannekoek' and Center for
High-Energy Astrophysics,  University of Amsterdam, Kruislaan 403,\\
1098 SJ Amsterdam, The Netherlands}
\date{}

\begin{abstract}

I discuss the energetics of synchrotron-emitting outflows,
increasingly found to be present in many different classes of X-ray
binary systems. It is shown that the outflow is likely to be comparable
in power to the integrated X-ray luminosity, traditionally taken to be
an indicator of the global mass-transfer rate. This is especially
found to be the case in the (low/)hard states of black hole candidate
systems. I conclude that jets are extremely important, energetically
and dynamically, for the accretion process in the majority of known
X-ray binary systems.

\end{abstract}

\end{opening}           

\section{Introduction -- the ubiquity of jets from X-ray binaries}  

Radio emission as a property of X-ray binary systems (XRBs) has
developed in recent years from being a rare and unusual facet of their
broadband spectrum to being recognised as a ubiquitous property of
several classes of XRBs.  In the early 1990s, it was still (almost)
possible to discuss individually radio-emitting X-ray binaries
(e.g. Hjellming \& Han 1995), and numbers of systems in which
collimated outflows, or jets, were directly observed could be counted
on the fingers of one hand.  As the 1990s drew on, more and more of
these systems became resolved into such jets, and in the middle of the
decade the apparent superluminal motions observed from GRS 1915+105
(Mirabel \& Rodriguez 1994; see also Fender et al. 1999) and GRO
J1655-40 (Tingay et al. 1995; Hjellming \& Rupen 1995) established
beyond doubt that XRBs could accelerate powerful flows to bulk
velocities in excess of $0.9c$. By 2000 we are beginning to realise
that jets may be rather more ubiquitous than previously imagined
(e.g. Mirabel \& Rodriguez 1999; Fender \& Hendry 2000; Fender 2000),
and in Figs 1(a),(b) I summarise our (or at least my) current
understanding of the empirical relation of radio emission to X-ray
`state' of the neutron star (NS) and black hole candidate (BHC)
systems respectively. In addition to the relations indicated in these
figures, it seems that discrete, radio-emitting ejection events (in
which the radio spectrum rapidly becomes optically thin) are
associated with state transitions (including outbursts of transients).

\begin{figure}
\centering\epsfig{figure=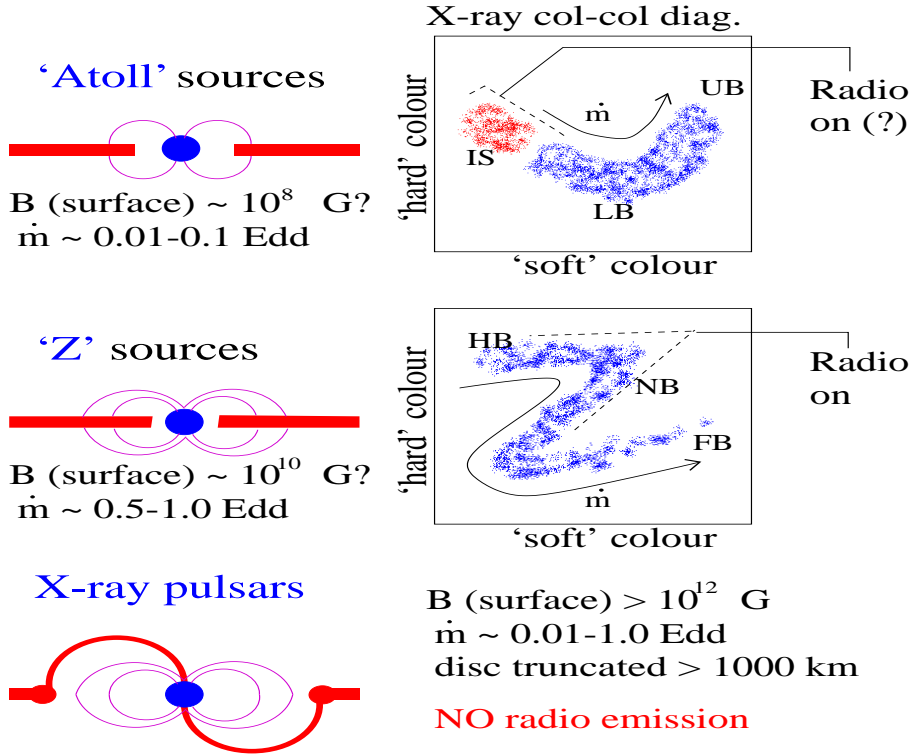,width=12cm,height=10cm,angle=0}
\caption{(a): A qualititative sketch of the relation of jets to
accretion in the three `types' of neutron star XRB. In the low-field
`Atoll' sources the accretion rate is believed to be $<10$\%
Eddington, except possibly during rare transient outbursts (e.g. Aql
X-1); the evidence is marginal so far but it appears such sources are
`radio on' when in the `Island State' (IS) in the X-ray colour-colour
diagram (CD). The Z sources are believed to be accreting at a much
higher rate, near Eddington, and are `radio on' when on the
`Horizontal Branch' and maybe also, at a lower level, the `Normal
Branch' in the CD. Note that for both Atoll and Z sources the
estimates of surface magnetic field are very uncertain. Finally, in
the high-field X-ray pulsars no radio synchrotron emission has ever
been detected; possibly this is due to truncation of the accretion
disc a long way from the neutron star.}
\end{figure}

\begin{figure}
\setcounter{figure}{0}
\centering\epsfig{figure=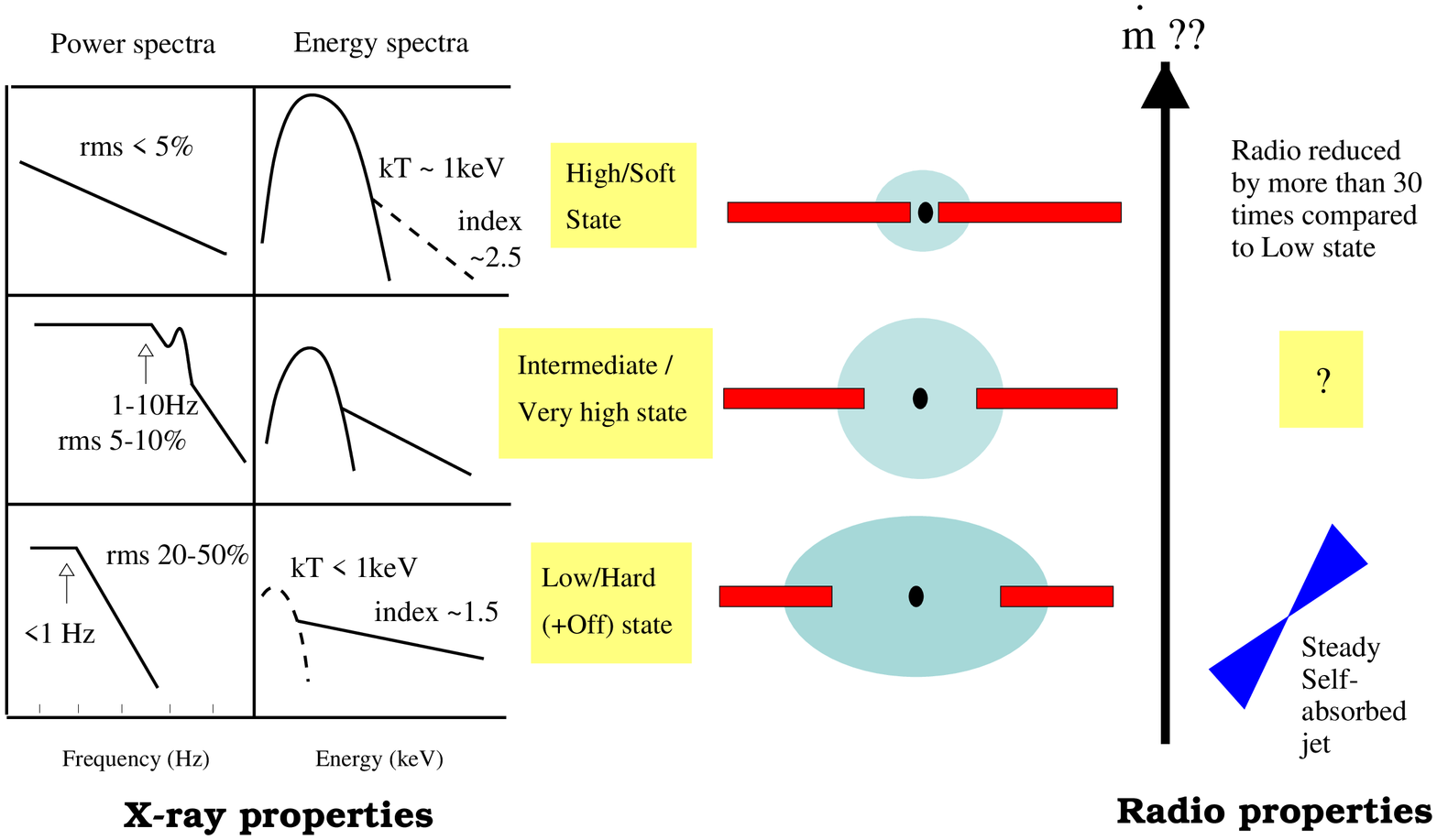,width=10cm,height=12cm,angle=270}
\caption{(b): The qualitative relation of radio emission to X-ray
`state' in BHC XRBs: the low/hard state is found to produce a steady,
flat- (or inverted-)spectrum jet, the soft state produces no
detectable radio emission. The relation of $\dot{m}$ to these states
is not certain (e.g. Homan et al. 2001); nor are the radio
characteristics 
of the relatively rare `Intermediate/Very High' states well determined.
}
\end{figure}

While some areas of the coupling between accretion flow and jet are
still empirically uncertain (the Atoll sources, the Intermediate/Very
High state for BHCs), it seems that all XRBs except the
high-field X-ray pulsars will, under the right conditions, produce a
synchrotron-emitting jet. It is therefore an important question to
address the significance of this jet, energetically and dynamically,
for the process of accretion onto compact objects as a whole. One
thing obvious from inspection of Figs 1(a),(b) is the apparent
anti-correlation between the mass accretion rate, $\dot{m}$,
traditionally estimated from X-ray studies alone, and the presence of
a jet, in both NS and BHC systems (see also Belloni, Migliari \&
Fender 2000 and Homan et al. 2001 for further discussion).

\section{Energetics from `equipartition'}

\begin{figure}
\centering\epsfig{figure=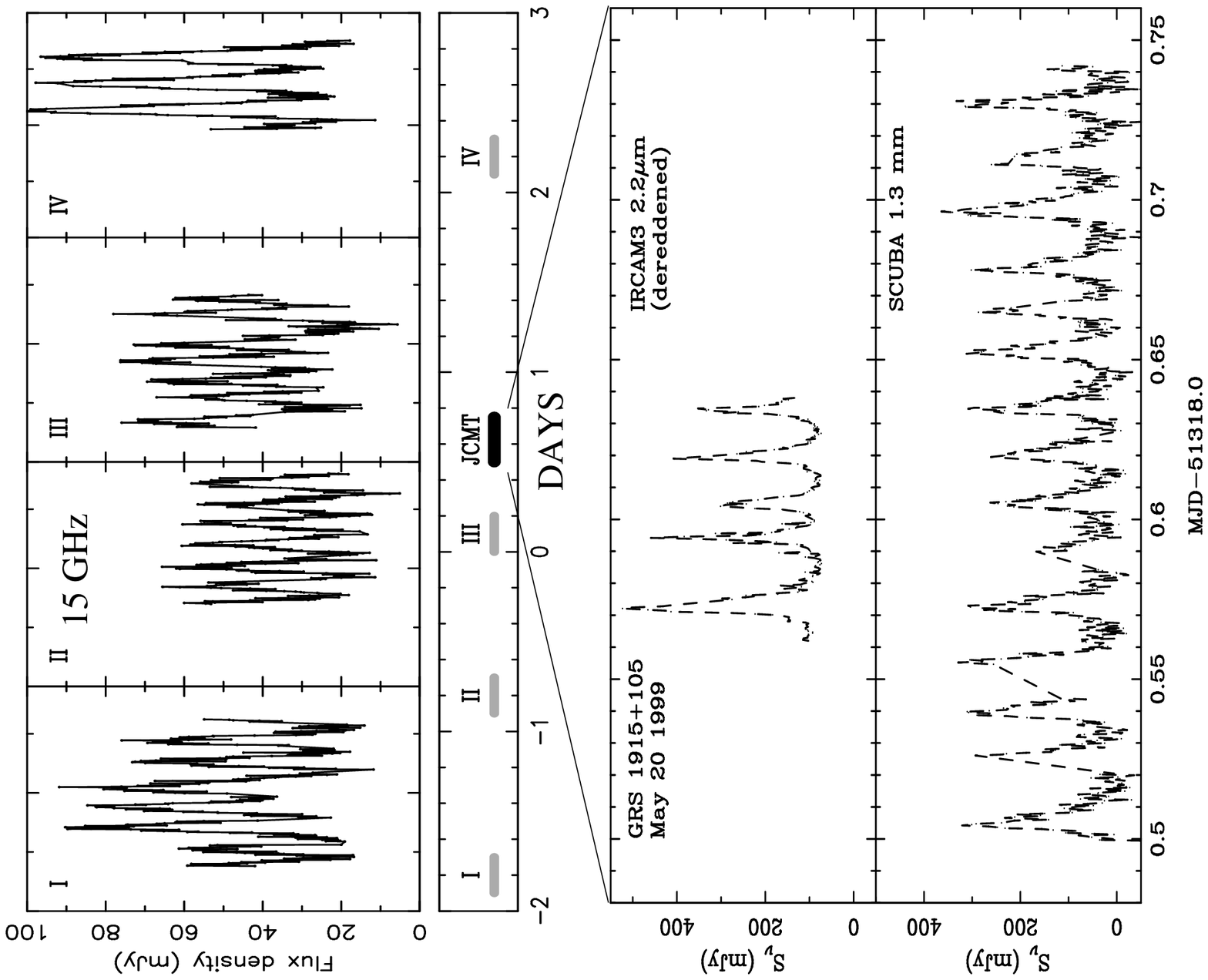,width=10cm,height=11cm,angle=270}
\caption{
Simultaneous mm and infrared observations of synchrotron
oscillations GRS 1915+105 (lower panel). Each event is believed to be
associated with a discrete ejection, and the observations came in the
middle of a period of more than a week of continuous oscillations as
observed at 15 GHz (top panel); see Fender \& Pooley (2000).
Note that each of the four 15 GHz observations lasts $\sim 0.1$ hr.
}
\end{figure}

In cases where it is possible to associate a given synchrotron
luminosity with a given volume, the minimum energy associated with the
event can be estimated assuming approximate equipartition. In this
situtation, the energy associated with the magnetic field is
approximately
equal to that associated with the electrons, and there is a
corresponding equipartition magnetic field, B$_{\rm eq}$.  Below the
equipartition field, the energy in particles ($\propto {\rm
B}^{-3/2}$) dominates; above equipartition, the energy in the field
($\propto {\rm B}^{2}$) dominates. While there is little obvious
physical justification for assuming equipartition, it is useful in
that it represents the case of maximal radiative efficiency for a
synchrotron-emitting plasma.

One obvious way to associate luminosity and volume is by observing
discrete ejection events sufficiently well that the rise time (and
therefore maximum volume) and peak flux can be accurately measured.
In Fig 2 we see an example of a sequence of repeated discrete ejection
events, observed from radio through the mm band to the
near-infrared band. Full details of the observations, and the
application of equipartition estimations to these data, are found in
Fender \& Pooley (2000). The results of the equipartition calculations
are presented in Table I.  A more basic introduction to minimum energy
requirements via equipartition can be found in e.g. Longair (1994).

From the estimates in Table I, it is clear that even in the minimum
energy case, ie. a pair plasma with no significant bulk kinetic energy, the
power required for the jet is $3 \times 10^{38}$ erg s$^{-1}$. The
integrated X-ray luminosity (presumably {\em not} coming from the jet ?)
at this time was $\sim 10^{39}$ erg
s$^{-1}$, so we find that even in this case the ratio of jet power to
total X-ray luminosity, $P_{\rm J} / L_{\rm X} \sim 0.3$. If we
assume that the ejections share the bulk relativistic
motion of the major ejections, then it seems $P_{\rm J} > L_{\rm X}$.

Spencer (1996) presents a summary of energy estimates based on the
same application of equipartition to radio varibility, for five
XRB systems. In all cases the equipartition magnetic field lies in the
range 0.01--1 Gauss, lower than that derived for the events from GRS
1915+105 considered above; however this may not be unreasonable
considering that the peak emission from the longer events considered
by Spencer would have originated in larger volumes. In all the five
cases considered by Spencer (1996), the apparent jet power was at
least equal to the integrated X-ray luminosity of the system.

In cases where the radio structure can be clearly resolved, the volume
responsible for a certain luminosity can be estimated directly from
images, and again equipartition applied.  Spencer et al. (these
proceedings) apply this method directly to the steady, resolved jets
in Cygnus X-1, and Paredes et al. (2000) apply it to jets from the
unusual system LS 5039, finding equipartition magnetic fields of 0.3
and 0.2 Gauss respectively. Both works further conclude that the
energy associated with the resolved outflows is large, $>10^{39}$ erg; 
however a major uncertainty in estimating the power in the outflow is
associated with how continuously it must be generated.

\begin{table}
\begin{tabular}{rllllllll}
\hline
Case & $\Gamma$ & L(erg s$^{-1}$) & B$_{\rm eq}$(G) & $E_{\rm min}$(erg) & P (erg s$^{-1
})$ &  $\dot{M}_{\rm jet}$(g s$^{-1}$) & $\eta$ \\ 
\hline
e$^+$:e$^-$ & 1 
& $3 \times 10^{37}$ & 40 & $4 \times 10^{41}$ & $3 \times
10^{38}$ & -- & $<0.1$ \\
e$^+$:e$^-$ & 5
& $4 \times 10^{39}$ & 115& $3 \times 10^{43}$ &
$3 \times 10^{40}$ & -- & $\sim 0.15$ \\
\hline
p$^+$:e$^{-}$ & 1 
& $3 \times 10^{37}$ & 40 & $4 \times 10^{41}$ &
$3 \times 10^{38}$ & $2 \times 10^{20}$ & $<0.1$\\
p$^+$:e$^{-}$ & 5 
& $5 \times 10^{39}$ & 115& $1 \times 10^{46}$ &
$8 \times 10^{42}$ & $4 \times 10^{21}$ & $<0.02$ \\
\hline
\end{tabular}
\caption{Calculation of (synchrotron) radiative luminosity,
equipartition magnetic field, total energy, jet power and mass-flow
rate for the oscillations reported here, given different physical
assumptions. $\Gamma$ is the bulk motion Lorentz factor; a filling
factor of unity is assumed, otherwise the equipartition magnetic field
would be high enough that radiative losses would be significant in the
infrared.  In these calculations a distance of 11 kpc and Doppler
factors for relativistic bulk motion which are the same as those
reported in Fender et al. (1999) are all assumed. Mass flow rate
$\dot{M}_{\rm jet}$ and jet power P are based upon one ejection every
20 min. For more details, see Fender \& Pooley (2000).}
\end{table}

\section{Steady jets -- energetics from radiative efficiency}

Without large amplitude variability, or directly resolved jets,
it is no longer possible to
associate a given luminosity with a certain volume, and we must apply
other arguments in order to estimate the total jet power. This is the
situation we face in trying to estimate the power in jets from BHCs
in the Low/Hard X-ray state, which produce a flat
spectrum in the radio band with no large-amplitude variability (Fender
2000). In this case we may estimate the total jet power by (a)
carefully measuring the extent of the synchrotron spectrum which it
produces, and (b) introducing a radiative efficiency, $\eta$, which is
the ratio of total to radiated power (in the jet's rest frame). From
this we can estimate the jet power as

\[
P_{\rm J} \sim L_{\rm J} \eta^{-1} F(\Gamma, i)
\]

where $L_{\rm J}$ is the total radiative luminosity of the jet, $\eta$
is the radiative efficiency, and $F(\Gamma, i)$ is a correction factor
for bulk relativistic motion with Lorentz factor $\Gamma$ and Doppler
factor $\delta$,
($F(\Gamma, i) \sim \Gamma \delta^{-3}$
-- see Fender 2000).

\begin{figure}
\centering\epsfig{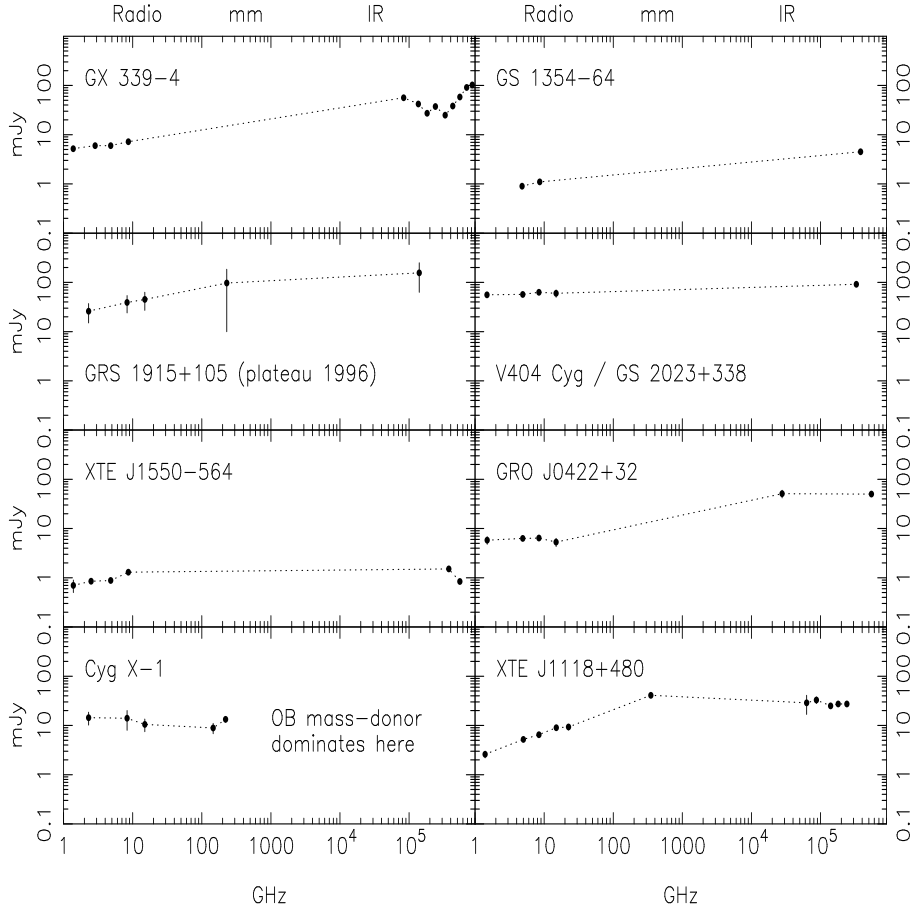}
\caption{Broadband radio--mm--infrared--optical spectra of eight black
hole candidate XRBs in hard X-ray states (apart from GRS 1915+105, all
of these can be considered to be the canonical `low/hard' X-ray
state). There seems to be clear evidence in all cases, where
observable (ie. not directly in Cyg X-1) for a continuation of the
radio--mm synchrotron spectrum to at least the near-infrared band.
Data and/or references for all systems in Fender (2000), except
GX 339-4 (infrared--optical data from Corbel \& Fender 2000),
XTE J1550-564 (Corbel et al. 2000) and XTE J1118+480 (Fender et al. 2000).
}
\end{figure}

\subsection{Spectral extent and radiative luminosity}

Starting from the reasonable assumption that all the emission observed
in the radio band is synchrotron in origin, we can try to see how far
this spectrum extends to other wavelengths. Firstly, it should be made
clear that most systems have not been observed at $\nu < 1$ GHz, and
while some low-frequency turnovers have occasionally been observed,
there are no reported cases of a complete cut-off to the synchrotron
emission at low radio frequencies. In any case, while a low-frequency
cut-off is important for estimating the mass of the ejecta in the (by
no means certain) case that there is a proton for each emitting
electron, the radiative luminosity is dominated by the high-frequency
extent of the synchrotron spectrum. Since the spectrum is unlikely to
have a spectral index $\alpha < -1$ (where $S_{\nu} \propto
\nu^{\alpha}$), then the most important observation is simply
determination of the highest frequency at which synchrotron emission
is observed.

In Fig 3 the broadband radio--optical spectra of eight XRB
systems in (low/)hard X-ray states is presented; in all cases the
spectra hint at an extension of the synchrotron spectrum to the
near-infrared or even optical bands. There is a large amount of
additional evidence for such a high-frequency extent to the spectrum,
each piece of which may be considered `circumstantial' but when
gathered together presents a compelling case (a lot of this evidence
is presented in Fender 2000). Note also that observations of GRS
1915+105 (admittedly not a `typical' system!), such as those shown in
Fig 2, unambiguously establish that the synchrotron spectrum can extend
to the near-infrared band.

Our best recent case is the low/hard state transient XTE J1118+480,
which clearly shows excess emission at near-infrared and probably also
optical wavelengths (Hynes et al. 2000) and whose radio spectrum
smoothly connects to a sub-mm detection at 850 $\mu$m (Fender et
al. 2000). In Fender et al. (2000) it is argued that in this case the
synchrotron radiative luminosity is already $\geq 1$\% of the
bolometric X-ray luminosity. How important the total jet power is then
depends on our estimates for the radiative efficiency, $\eta$.

\subsection{Radiative efficiency}

In Fender \& Pooley (2000) a detailed study was made of the power
required to maintain the repeated ejection events shown in Fig 2.
In Table I, we also tabulate this radiative efficiency for the
different cases considered; its maximum value is $\sim 0.15$. 
In one of the original works on the power in steady jets, Blandford \&
K\"onigl (1979) determined that $\eta \sim \frac{1}{2}k_e
(1+\frac{2}{3}k_e \Lambda)^{-1}$, where $k_e \leq 1$ and $\Lambda =
\ln (\gamma_{\rm e,max})/(\gamma_{\rm e,min})$, $\gamma_{\rm
e,max/min}$ being the upper and lower bounds of the electron energy
spectrum respectively. Since it seems likely that $(\gamma_{\rm
e,max}/(\gamma_{\rm e,min})$ is in the range $10^{2}$--$10^{4}$, then
$\eta < 0.15$. Celotti \& Ghisellini (2001) find a comparable value
for the radiative efficiency of AGN jets. 

\section{Comparison of jet power to accretion luminosity}

The best-studied cases of clearly resolved (in time) ejection events,
such as those presented in Fig 2, seem to require that the ratio of
jet power to X-ray luminosity, $P_{\rm J} / L_{\rm X} >
0.3$. A more general study of large outbursts from five different
systems shows that, via the simple equipartition arguments outlined
above,  $P_{\rm J} / L_{\rm X} > 1$ (Spencer 1996). Other applications
of equipartition to XRB ejection events exist, and in all cases it
seems that they imply the jet power is extremely significant.

Using a different, but not altogther unrelated, method of measuring
the high-frequency extent of the synchrotron spectrum we find that for
the BHC transient XTE J1118+480, $P_{\rm J} / L_{\rm X} >
0.01 \eta^{-1}$, where $\eta < 0.15$ for BHC systems in the canonical
low/hard X-ray state. Since it seems that the relation of X-ray to
radio flux in this state is comparable for all the systems (ie. the
similarities in the broadband spectra of the sources presented in Fig
3 extend to at least the X-ray band), then it seems reasonable to
conclude that $P_{\rm J} / L_{\rm X} \geq 0.1$ for all BHC XRBs in the
low/hard X-ray state.  Again it seems that the jet must be highly
significant for the process of accretion. 

\section{Conclusions}

Two different (but not entirely independent) approaches have revealed
that a large fraction of the entire luminosity (radiative and
mechanical) of some {\em classes} (this is important -- we are no
longer talking about individual `oddballs') of X-ray binaries may be
carried by the jet.  This fraction is likely to be $> 0.1$ and may be
$>0.5$, ie. the jet may even be the {\em dominant} power output
channel. Since it is clear from studies of e.g. GRS 1915+105 that the
disc--jet coupling is occurring in the inner few 100 km of the
accretion flow, the extraction of such a large amount of energy (and
presumably angular momentum) must have a dramatic effect on the
accretion flow close to the compact object.

Based on this, it seems that models for accretion in X-ray binaries
which at least parameterise such an outflow are needed. They are not
entirely non-existent -- for GRS 1915+105 (again!) both Nayakshin,
Rappaport \& Melia (2000) and Janiuk, Czerny \& Siemiginowska (2000)
have begun to consider the effect of a moderately powerful outflow on
the accretion process. A more radical step has been taken by Markoff,
Falcke \& Fender (2000; see also these proceedings) in which almost
the entire broadband spectrum of the BHC low state (excepting thermal
emission in the optical -- UV range) is explained by jet emission, and
in which it is suggested that the X-ray power-law may even be {\em
synchrotron} emission. This would be contrary to the generally
accepted view that the X-ray power law in the low state arises in
Comptonisation of `seed' photons in a hot, thermal, corona. It is
further worth noting that once mass-loss via jets is considered, and
the empirical connections between X-ray `state' and jet formation, as
outlined in Figs 1(a),(b) recognised, then it is no longer clear that
the global mass transfer rate
can be inferred from X-ray observations
alone.

\acknowledgements

I would like to thank many of the participants at the Granada workshop
for stimulating conversations, and Annalisa Celotti and Kieran O'Brien
for reading through a draft of this work. In particular I would like
to acknowledge many relevant conversations, over several years, with
Ralph Spencer, Guy Pooley and the late Bob Hjellming.

\theendnotes

\end{article}
\end{document}